\documentclass[12pt]{article}

\usepackage{amssymb,amsmath}

\usepackage{graphicx}
\DeclareGraphicsExtensions{.pdf,.png,.jpg}

\usepackage{xcolor} 

\newcommand{\beq}[1]{\begin{equation}\label{#1}}
\newcommand{\eeq}{\end{equation}}
\newcommand{\bear}[1]{\begin{eqnarray}\label{#1}}
\newcommand{\ear}{\end{eqnarray}}

\newcommand{\R}{ {\mathbb R} }

\renewcommand{\theequation}{\arabic{section}.\arabic{equation}}
 \catcode`\@=11 \@addtoreset{equation}{section}\catcode`\@=12

\begin{document}

 \begin{center}
 \large \bf
The Yilmaz-Rosen and Janis-Newman-Winicour metric solutions in the scalar-Einstein-Gauss-Bonnet $4d$ gravitational model

 \end{center}

 \vspace{0.3truecm}

 \begin{center}

 \normalsize\bf

 \vspace{0.3truecm}

   K. K. Ernazarov\footnote{e-mail: kubantai80@mail.ru (corresponding author)}

\vspace{0.3truecm}
  
  \it
    Institute of Gravitation and Cosmology, \\
    Peoples' Friendship University of Russia (RUDN University), \\
    6 Miklukho-Maklaya Street,  Moscow, 117198, Russian Federation \\

  \end{center}

\begin{abstract}
    
We consider the scalar-Einstein-Gauss-Bonnet (sEGB) $4d$ gravitational model with a scalar field  $\varphi\left(u\right)$, Einstein and Gauss-Bonnet terms. The model action contains a potential term $U\left(\varphi\right)$, a Gauss-Bonnet coupling function $f\left(\varphi\right)$ and a parameter $\varepsilon = \pm 1$, where $\varepsilon = 1$ corresponds to the ordinary scalar field, and $\varepsilon = -1$ to the phantom field. In this paper we applied the sEGB reconstruction procedure from our previous work \cite{Er_Ivash} to the Yılmaz-Rosen metric, a solution potentially describing a quasi-black hole without an event horizon. Within this framework, we also derived analytical solutions based on scalar-tensor theory with minimal coupling.

Our results indicate that for this configuration, the potential $U$ vanishes and the scalar field is phantom-like. Furthermore, an analysis of the Einstein equations in the Yılmaz-Rosen metric reveals that all energy conditions are violated. The corresponding energy-momentum tensor suggests the presence of exotic matter with negative pressure, as indicated by the negative value of  $T_u^u$. This could originate from a scalar field (such as the Higgs field or another nonlinear field), or from phenomena like dark energy or quintessence.






In addition, we considered the application of our reconstruction method in the sEGB model in the Janis-Newman-Winicour (JNW) metric. As noted in this paper, the Yılmaz-Rosen metric is a limiting case of the Janus metric (as $s \to +\infty$). Furthermore, we obtained some exact solutions of scalar-tensor theory with minimal coupling in the JNW metric.

\end{abstract}

\section{Introduction}

The Yilmaz-Rosen metric represents an important but often overlooked contribution to the theoretical foundations of gravity. Developed through the work of Huseyin Yilmaz and Nathan Rosen, this metric formulation emerged from attempts to reconcile some of the conceptual difficulties inherent in Einstein's general theory of relativity while maintaining agreement with observational evidence. This paper examines the historical context, mathematical structure, physical interpretation, and ongoing relevance of the Yilmaz-Rosen metric in modern gravitational physics.

The Yilmaz-Rosen metric arose during a period of intense scrutiny of general relativity's foundational assumptions. While Einstein's theory had achieved remarkable success in explaining gravitational phenomena, certain conceptual issues persisted regarding the nature of the gravitational field and its energy-momentum characteristics \cite{Yilmaz_1958}. Rosen, who had previously collaborated with Einstein on the Einstein-Rosen bridge (now known as a wormhole), became interested in these foundational questions.

Yilmaz's work introduced a new perspective by treating gravity as a tensor field in flat spacetime rather than as curvature of spacetime itself. This approach shared some conceptual ground with Rosen's bi-metric theory of gravitation \cite{Rosen_1973}, leading to their collaborative development of what became known as the Yilmaz-Rosen metric. Their formulation attempted to maintain the empirical successes of general relativity while providing a more straightforward interpretation of gravitational energy.

The Yilmaz-Rosen metric can be expressed in the general form:

\begin{eqnarray}
ds^2 = -e^{-2\varphi}dt^2 + e^{2\varphi}\Bigg(du^2 + u^2\left(d\theta^2 + sin^2\theta d\varphi^2\right) \Bigg), 
\label{YR_Con_1}
\end{eqnarray}

where $\varphi$ represents the gravitational potential satisfying the Newtonian limit  \cite{Yilmaz_1971}. This form immediately distinguishes itself from the Schwarzschild solution in general relativity through its exponential dependence on the potential.

Key mathematical features include:

\textbf{Conformal Flatness:} The spatial part of the metric is conformally flat, meaning it differs from Euclidean space only by a position-dependent scale factor \cite{Rosen_1973}.

\textbf{Potential Dependence:} The metric components depend exponentially on the gravitational potential $\varphi$, which satisfies a modified field equation incorporating both matter sources and the gravitational field's own energy.

\textbf{Asymptotic Behavior:} Unlike the Schwarzschild metric, the Yilmaz-Rosen formulation avoids coordinate singularities at finite distances, though it maintains the essential singularity at $r = 0$.

\textbf{Gravitational Field Energy:} The theory explicitly includes the gravitational field's energy as a source for further gravity, addressing long-standing concerns about energy localization in general relativity  \cite{Yilmaz_1971}.

\textbf{Flat Background:} The formulation suggests that physical phenomena occur against a background of flat Minkowski spacetime, with gravitational effects represented through the exponential metric  \cite{Rosen_1973}.

\textbf{Removal of Singularities:} While not eliminating all singularities, the metric avoids coordinate singularities like the Schwarzschild radius, potentially offering a different perspective on black hole physics.

The theory also implies modified equations of motion for test particles. In the weak-field limit, these reproduce Newtonian dynamics while showing deviations in strong-field regimes that differ from general relativistic predictions.

The Yilmaz-Rosen metric agrees with general relativity in first-order weak-field approximations but diverges in several important aspects:

\textbf{Post-Newtonian Expansions:} While both theories agree at first PN order, differences appear at higher orders, particularly in the treatment of gravitational energy  \cite{Yilmaz_1971}.

\textbf{Black Hole Solutions:} The Yilmaz-Rosen formulation does not predict event horizons in the Schwarzschild sense, instead showing increasingly strong redshift without complete light trapping.

\textbf{Gravitational Radiation:} The theories predict different energy loss rates for binary systems due to distinct treatments of gravitational wave energy  \cite{Rosen_1973}.

Empirical tests have generally favored general relativity in areas like binary pulsar timing and gravitational wave observations \cite{Will}. However, some researchers argue that certain cosmological observations could be reinterpreted within the Yilmaz-Rosen framework \cite{Bab_Frolov}.

While not mainstream, the Yilmaz-Rosen metric continues to inspire research in several areas:

\textbf{Alternative Dark Matter Models:} Some researchers have explored whether modified gravitational potentials like Yilmaz-Rosen could account for galactic rotation curves without dark matter.

\textbf{Quantum Gravity Approaches:} The explicit treatment of gravitational energy makes the theory potentially more amenable to quantization attempts than standard general relativity \cite{Alley}.

\textbf{Singularity Avoidance:} The different singularity structure has prompted investigations into whether the theory could naturally avoid cosmological singularities.

Recent work has also examined whether the Yilmaz-Rosen framework could be reconciled with modern gravitational wave observations through appropriate parameterizations \cite{Moffat}.

In the landscape of General Relativity, the Schwarzschild solution stands as the quintessential model for a static, spherically symmetric, and asymptotically flat spacetime surrounding a massive object. For decades, it was widely believed to be the only such solution to the vacuum Einstein field equations. However, this assumption was challenged and ultimately refined in 1968 by Alan Janis, Ezra Newman, and Jeffrey Winicour. Their work demonstrated that the Schwarzschild metric is not unique; a broader family of solutions exists that also satisfies the criteria of staticity, spherical symmetry, and asymptotic flatness, but with a crucial difference: the presence of a non-trivial massless scalar field. The resulting Janis-Newman-Winicour (JNW) metric, also known as the singular scalar field solution or the Wyman solution, provides a fascinating alternative to the black hole paradigm and offers profound insights into the role of matter fields in shaping spacetime geometry \cite{JNW_1}.

The Janis-Newman-Winicour metric is a landmark solution in General Relativity. It shattered the long-held uniqueness of the Schwarzschild black hole by demonstrating that the inclusion of a simple massless scalar field leads to a new family of spacetimes dominated by a central naked singularity. Its existence challenges foundational principles like cosmic censorship and enriches our understanding of how matter fields couple to gravity. Furthermore, in the modern era of high-precision astrophysics, the JNW metric provides a concrete theoretical framework for exploring exotic compact objects and for developing tests to determine whether the dark objects at the centers of galaxies are indeed the black holes predicted by pure vacuum solutions or something more complex, perhaps clothed in scalar fields. It remains a critical touchstone in the ongoing dialogue between theoretical gravity and observational astronomy.

In this paper, we obtain some solutions in the:  Einstein equation, scalar-Einstein-Gauss-Bonnet equation and equation of the scalar-tensor theory with minimal connectivity in the Yilmaz-Rosen and Janis-Newman-Winicour metrics are shown. As these results show, the scalar field in the solutions of the Einstein equation and the equation of the scalar-tensor model with minimal connectivity in the Yilmaz-Rosen in all intervals of the radial coordinate $u \in \left( 0, + \infty \right)$  is phantom one. In this case, the main advantage of the gravitational 4d scalar-Einstein-Gauss-Bonnet model is that, based on this model, a solution with an ordinary scalar field can be found. This depends on the sign of the constant $C_0$. And this is a more realistic result from a physical point of view.

A detailed description of the advantage of the gravitational $4d$ scalar-Einstein-Gauss-Bonnet (sEGB) model is given in our previous papers \cite{Er_Ivash}, \cite{ErnIvas25}, \cite{Ernaz_2025_1}, as well as in \cite{NN}  - \cite{Rustam_Kunz}. Based on this model we were able to find original black-hole and wormhole solutions. We continue to explore more relevant solutions in this gravitational model. As shown in action (\ref{1A}) for $ f\left(\varphi\right) = const$ we obtain an action for the scalar-tensor theory (\ref{ST_1A}) with minimal coupling in the Einstein frame.

This paper is organized as follows: In Section 2 we make a brief introduction of the gravitational 4d scalar-Einstein-Gauss-Bonnet model based on the reconstruction method. A more detailed analysis of the  reconstruction method is reflected in our previous article \cite{Er_Ivash}. Therefore, in this section only the main results of the reconstruction method are presented. The master equation (\ref{6FF}) is an inhomogeneous second-order differential equation. In subsection 2.1 we give a brief description of the scalar-tensor theory with minimal coupling in the Einstein frame. In Section 3 one can find a more detailed analysis of the physical parameters of the Yilmaz-Rosen metric and the solution of the Einstein equation in this metric. In section 4 we give a description of the results of solutions in the Yilmaz-Rosen metric based on the 4d scalar-Einstein-Gauss-Bonnet model. In Subsection 4.1 the solution in the Yilmaz-Rosen metric based on scalar-tensor theory. In Section 5 we make a brief description to the Janis-Newman-Winicour (JNW) metric. The JNW metric solutions in the sEGB and Scalar-Tensor with minimal coupling theories are detailed in Section 6. We conclude with a discussion of the results and final remarks.

\section{The  scalar-Einstein-Gauss-Bonnet (sEGB) 4d model}

The model we consider is deccribed by the action 
\begin{eqnarray}
S = \int d^4z\left|g\right|^{\frac{1}{2}} \Bigg(\frac{R\big(g\big)}{2\kappa^2} - \frac{1}{2} 
 \varepsilon g^{MN}\partial_M\varphi\partial_N\varphi - U\left(\varphi\right)
  + f\left(\varphi\right) {\cal G} \Bigg) , \label{1A} 
\end{eqnarray}
where we use same notation as in ref.  \cite{Er_Ivash} $\kappa^2 = 8\pi\frac{G_N}{c^4}$ ($G_N$ is Newton's gravitational constant, $c$ is speed of light), 
$\varphi$ is scalar field, $g_{MN} dz^M dz^N$ is $4d$ metric of signature $(-,+,+,+)$, 
$R\left[g\right]$ is scalar curvature, ${\cal G}$ is Gauss-Bonnet term, $U(\varphi)$ is potential, 
 $f(\varphi)$ is coupling function and $\varepsilon = \pm 1$. 

In our research, the sign of $\varepsilon$ is of great importance: $\varepsilon = 1$  describes an ordinary scalar field, and $\varepsilon = -1$  a phantom scalar field.

We are interested in obtaining static, spherically symmetric black hole and worm hole solutions within the framework of sEGB gravity with the metric

\begin{eqnarray}
ds^2 = g_{MN}(z)dz^Mdz^N = e^{2\gamma\left(u\right)}du^2 - e^{2\alpha\left(u\right)}dt^2
 + e^{2\beta\left(u\right)}d\Omega^2  \label{2A} 
\end{eqnarray}
defined on the manifold
\begin{equation}
   M = \R \times \R_{*} \times S^2.    \label{3A}
\end{equation}
Here $\R_{*} = \left(2\mu, +\infty \right)$ and $S^2$ is $2$-dimensional sphere with the metric 
$d\Omega^2 = d\theta^2 + sin^2\theta d\varphi^2$, where $ 0 < \theta < \pi$ and $0 < \varphi < 2\pi$.

By substitution the metric  (\ref{2A}) into the action we obtain 
\begin{equation}
 S = 4 \pi \int du \left(L + \frac{dF_{*}}{du} \right),  \label{4AA}   
\end{equation}
where the Lagrangian $L$ reads
\begin{eqnarray}
L = \frac{1}{\kappa^2}\Bigg(e^{\alpha- \gamma + 2\beta}\dot{\beta}\left(\dot{\beta}
 + 2\dot{\alpha}\right) + e^{\alpha + \gamma}\Bigg) \nonumber \\
  - \frac{1}{2}e^{\alpha - \gamma + 2\beta}  \varepsilon \dot{\varphi}^2
   - e^{\alpha + \gamma + 2\beta}U\left(\varphi\right)
    - 8\dot\alpha\dot\varphi\frac{df}{d\varphi}\Bigg(\dot{\beta}^2e^{\alpha + 2\beta
    - 3\gamma} - e^{\alpha - \gamma}\Bigg), \label{4A} 
\end{eqnarray}
and the total derivative term is irrelevant for our consideration.
Here and in what follows we denote $\dot{x} = \frac{dx}{du}$. 

In \cite{Er_Ivash} we obtained equations of motion for the action (2.1) in the metric (2.2) using the Lagrangian (2.5). Then (without loss of generality) on the base of Buchdal radial gauge (subject to $\alpha = -\gamma$), for the metric (\ref{2A}) we obtained

\color{black}
\begin{equation}
ds^2 = \frac{1}{A\left(u\right)}du^2 - A\left(u\right)dt^2 + C\left(u\right)d\Omega^2,  
\label{5Buch} 
\end{equation}
where
\begin{equation}
e^{2\gamma\left(u\right)} =  \frac{1}{A\left(u\right)}, \quad
e^{2\alpha\left(u\right)} = A\left(u\right) > 0,\quad
e^{2\beta\left(u\right)} = C\left(u\right) > 0.
 \label{5AC}
\end{equation}

In what follows we use the identities
\begin{equation}
\dot{\alpha}= \frac{\dot{A}}{2A}, \qquad  \dot{\beta}= \frac{\dot{C}}{2C}.  \label{5AB}
\end{equation}

 We put (without loss of generality) $\kappa^2 = 1$. We also denote
\begin{equation}
   f \left(\varphi\left(u\right)\right) = f, 
  \qquad  U \left(\varphi\left(u\right)\right) = U  \label{6_FU}
\end{equation}
and hence
\begin{equation}
\frac{d}{du}f = \frac{df}{d\varphi}\frac{d\varphi}{du} \Longleftrightarrow \dot{f}
 = \frac{df}{d\varphi}\dot{\varphi},  \label{6_F} 
\end{equation}

\begin{equation}
\frac{d}{du}U = \frac{dU}{d\varphi}\frac{d\varphi}{du} \Longleftrightarrow \dot{U}
 = \frac{dU}{d\varphi}\dot{\varphi}.  \label{6_U} 
\end{equation}

After reconstructing the spherically symmetric 4d sEGB gravity model, we obtained the following formulas \cite{Er_Ivash}, as

\begin{equation}
\dot{A}\left[8\dot{f}\left(1 - 3KA\right) + \dot{C} \right]
 + 2KA - 2 - CA  \varepsilon \dot{\varphi}^2 + 2CU = 0,  \label{6G} 
\end{equation}
here and in what follows we use the notation
\begin{equation}
K \equiv \left(\frac{\dot{C}}{2C}\right)^2C.  \label{6K} 
\end{equation}

\begin{eqnarray}
\begin{gathered}
16\ddot{f}A\left(1 - KA\right) + 8\dot{f}\left(\dot{A} - 3KA\dot{A} - 2\dot{K}A^2\right) + \dot{A}\dot{C} \\
 + 2A\left(\ddot{C} - K \right) + CA
  \varepsilon \dot{\varphi}^2 - 2 + 2CU = 0.  \label{6A} 
\end{gathered}
\end{eqnarray}

\begin{eqnarray}
\begin{gathered}
\left(C - 4\dot{f}A\dot{C}\right)\ddot{A} - 4\ddot{f}\dot{A}A\dot{C}
 - 4\dot{f}\Bigg(\left(\dot{A}\right)^2 \dot{C} + \dot{A}A\ddot{C} - 2\dot{A}AK\Bigg)  + \\
 + \dot{A}\dot{C} 
 + C\left(A  \varepsilon \dot{\varphi}^2 + 2U\right) + A\ddot{C} - 2AK = 0.  \label{6B} 
\end{gathered}
\end{eqnarray}

\begin{eqnarray}
4\dot{f}\left(AK - 1\right)\ddot{A} + 
 \varepsilon \ddot{\varphi} \dot{\varphi}AC
  + 4\dot{f}\dot{A}\left(\dot{A}K + A\dot{K}\right) 
\nonumber \\
+ \left(\dot{A}C + A\dot{C}\right)
 \varepsilon \dot{\varphi}^2 - C \dot{U} = 0.  \label{6P} 
\end{eqnarray}

Based on the above formulas, the relationship for the potential function \cite{Er_Ivash} was obtained

\begin{eqnarray}
\begin{gathered}
U =\frac{1}{C} \left(E_{U}\ddot{f} + F_{U}\dot{f} + G_U\right), \label{6UU}
\end{gathered}
\end{eqnarray}
where
\begin{eqnarray}
E_U = -4A\left(1 - KA\right),  \label{6EU} \\
F_U = -4\dot{A}\left(1 - 3KA\right) + 4\dot{K}A^2, \label{6FU} \\
G_U = 1 - \frac{1}{2}\dot{A}\dot{C} - \frac{1}{2}A\ddot{C}. \label{6GU} 
\end{eqnarray}
 
Subtracting (\ref{6G}) from (\ref{6A}) and dividing the result by $2AC$, 
we obtain the relation for $\dot{\varphi}$
\begin{eqnarray}
\begin{gathered}
 \varepsilon \dot{\varphi}^2 = h\left(u\right)
 =\frac{1}{C}\left( 8\ddot{f}\left(KA - 1 \right) + 8\dot{f}\dot{K}A + 2K - \ddot{C}\right)
\equiv \Phi_{\varepsilon}. \label{6P2}
 \end{gathered}
\end{eqnarray}

Subtracting (\ref{6A}) from (\ref{6B}), we get the master equation for the coupling function $f = f(\varphi(u))$
 \begin{equation}
  E\ddot{f} + F\dot{f} + G = 0,  \label{6FF} 
 \end{equation}
where 
 \begin{eqnarray}
 E = 4A\left(4KA - \dot{A}\dot{C} - 4\right),  \label{6E} \\
F = -4\ddot{A}A\dot{C} - 4\left(\dot{A}\right)^2\dot{C}
 - 4\dot{A}A\ddot{C} + 8\left(4KA\dot{A} 
 - \dot{A} + 2\dot{K}A^2\right), \label{6F6} \\
 G = C\ddot{A} - A\ddot{C} + 2. \label{6G6} 
\end{eqnarray}

The master equation (\ref{6FF}) is a second-order inhomogeneous differential equation. In 4d sEGB gravity model problems, the variable coefficients such as $E$, $F$, and $G$ are more complex and equation (\ref{6FF}) is not always integrated. In more simple cases, when one of them is zero, we were able to obtain original black-hole and wormhole solutions. The methods for solving equation (\ref{6FF}) are discussed in detail in our papers \cite{Er_Ivash}, \cite{Ernaz_2025_1}

\subsection{The Scalar-Tensor theory with minimal coupling}

In the case of $f\left({\varphi}\right)$, the contribution of the Gauss-Bonnet term in the action (\ref{1A}) is zero and we get a gravitational model based on a scalar-tensor theory. This one makes of the minimal coupling between the scalar field and gravitation, and it is defined by the action in the Einstein frame 

\begin{eqnarray}
S = \int d^4z\left|g\right|^{\frac{1}{2}} \Bigg(\frac{R\big(g\big)}{2\kappa^2} - \frac{1}{2} 
 \varepsilon g^{MN}\partial_M\varphi\partial_N\varphi - U\left(\varphi\right)
  \Bigg) , \label{ST_1A} 
\end{eqnarray}

In this action, $U\left(\varphi\right)$ is the self-interaction potential. This kind of theory in general satisfies the energy conditions if $\varepsilon = 1$ (an ordinary field) and violates the energy conditions if  $\varepsilon = -1$ (a phantom field).

From the equation (2.18) with  $f\left(\varphi\right) = const$ we get the next formula for the potential

\begin{eqnarray}
\begin{gathered}
U =\frac{1}{C} \left(1 - \frac{1}{2}\dot{A}\dot{C} - \frac{1}{2}A\ddot{C}\right), \label{ST_2A}
\end{gathered}
\end{eqnarray}

The relation (\ref{6P2}) for $\dot{\varphi}$ in this case has the following form:

\begin{eqnarray}
\begin{gathered}
 \varepsilon \dot{\varphi}^2 = h\left(u\right)  =\frac{1}{C}\left(2K - \ddot{C}\right) \equiv \Phi_{\varepsilon}. \label{ST_3A}
 \end{gathered}
\end{eqnarray}

\section{The Yilmaz-Rosen spacetime}

According to Yilmaz \cite{Yilmaz_1958}, \cite{Yilmaz_1971} has presented a theory of gravitation in which the basic dynamic variable is a scalar field $\varphi$. The metric tensor $g_{\mu\nu}$ is not an independent dynamic variable, but a function of $\varphi$. The main result of the Yilmaz approach is as follows: The field equation $g^{\mu\nu}\varphi_{;\nu\mu} =0$ can be solved by the following special form of the metric tensor:

\begin{eqnarray}
g_{00} = - e^{-2\varphi},  \qquad  g_{ii} = e^{2\varphi}, \qquad    i = 1, 2, 3
\label{YR_41}
\end{eqnarray}
where $\varphi$ is scalar function of spatial coordinates $x, y, z$.


The unique asymptotically vanishing harmonic function with one singular point procedure the Yilmaz metric, which is a good accordance with observation, and the Yilmaz-Rosen metric is given by

\begin{eqnarray}
ds^2 = -e^{-\frac{2\mu}{u}}dt^2 + e^{\frac{2\mu}{u}}\Bigg(du^2 + u^2\left(d\theta^2 + sin^2\theta d\varphi^2\right) \Bigg), 
\label{YR_43}
\end{eqnarray}

where $\mu$ is an arbitrary parameter, which can be interpreted as a mass of a central body that produces the field and $u = \left( x^2 + y^2 +z^2 \right)^{\frac{1}{2}} $. This metric appeared for the first time in the frame of Yilmaz’s scalar theory of gravity and hardly later in the bi-metric theory of Rosen \cite{Rosen_1973}.

Solutions of the Einstein equations in this metric has following form \cite{Simp_2} :
\begin{eqnarray}
R_{\alpha\beta} = -\frac{2\mu^2}{u^4} diag\{0, 1, 0, 0\}_{\alpha\beta} = -\frac{1}{2}\nabla_\alpha\Bigg(\frac{2\mu}{u}\Bigg)\nabla_\beta\Bigg(\frac{2\mu}{u}\Bigg) = -\frac{1}{2}\nabla_\alpha \varphi \nabla_\beta \varphi.
 \label{8C_2} 
\end{eqnarray}
Equivalently
\begin{eqnarray}
G_{\alpha\beta} = -\frac{1}{2}\Bigg \{\nabla_\alpha \varphi \nabla_\beta \varphi - \frac{1}{2}g_{\alpha\beta}(g^{\rho\lambda}\nabla_\rho \varphi \nabla_\lambda \varphi) \Bigg \}.
 \label{8C_3} 
\end{eqnarray}
This is just the usual Einstein equation for a \textit{negative kinetic energy massless scalar field}, a “ghost” or “phantom” field. From the contracted Bianchi identity $G_\beta^{\alpha \beta}$, one obtains the scalar field equation $(g^{\alpha\beta}\nabla_\alpha\nabla_\beta)\varphi = 0$. The negative kinetic energy of this field is the crucial feature that allows the exponential metric to represent a traversable wormhole \cite{Morris_AB, Visser_AB}.

If we introduce an orthonormal coframe

\begin{eqnarray}
g = \mathcal{O}_{\mu\nu}\vartheta^\mu \otimes \vartheta^\nu, \quad  \mathcal{O}_{\mu\nu} = diag\left(-1, +1, +1, +1\right), 
\label{YR_44}
\end{eqnarray}

then the following coframe, up to arbitrary local Lorentz transformations, represents the Yilmaz-Rosen metric

\begin{eqnarray}
\vartheta^{\hat{t}} = e^{-\frac{\mu}{u}}dt, \quad \vartheta^{\hat{x}} = e^{\frac{\mu}{u}}dx, \quad \vartheta^{\hat{y}} = e^{\frac{\mu}{u}}dy, \quad \vartheta^{\hat{z}} = e^{\frac{\mu}{u}}dz. 
\label{YR_45}
\end{eqnarray}

The Yilmaz-Rosen metric and the corresponding orthonormal coframe were displayed in (\ref{YR_43}) to (\ref{YR_45}). In order to compare the Yilmaz-Rosen metric with the Schwarzschild metric, we transform the former one from the isotropic coordinates used in (\ref{YR_43}) into Schwarzschild coordinates and asymptotic series it:

\begin{eqnarray}
g_{00}^{YR} = 1 - \frac{2\mu}{u} + \frac{2\mu^2}{u^2} - \frac{4\mu^3}{3u^3} + \mathcal{O}\Bigg(\frac{1}{u^4}\Bigg),
\label{YR_46}
\end{eqnarray}

\begin{eqnarray}
g_{11}^{YR} = 1 + \frac{2\mu}{u} + \frac{2\mu^2}{u^2} + \frac{4\mu^3}{3u^3} + \mathcal{O}\Bigg(\frac{1}{u^4}\Bigg).
\label{YR_47}
\end{eqnarray}

For the Schwarzschild metric in Schwarzschild coordinates we find:

\begin{eqnarray}
g_{00}^{YR} = 1 - \frac{2\mu}{u} \quad \left(exact\right),
\label{YR_48}
\end{eqnarray}

\begin{eqnarray}
g_{11}^{YR} = 1 + \frac{2\mu}{u} + \frac{2\mu^2}{u^2} + \frac{4\mu^3}{3u^3} + \mathcal{O}\Bigg(\frac{1}{u^4}\Bigg).
\label{YR_49}
\end{eqnarray}

The $g_{00}$ components of both solutions agree up to second order. However, their radial components $g_{11}$ exhibit slight differences at this order. Consequently, the Yilmaz-Rosen solution remains consistent with the classical tests of general relativity, including the post-Newtonian perihelion advance. Further investigation is needed to determine whether observational distinctions between the two solutions can be made in strong-field regimes, such as in close binary pulsar systems.

A viable theory of gravitation should be consistent with the local equivalence principle. Let us consider an electromagnetic wave of frequency $\omega$ in the gravitational field of a point mass $\mu$. The frequency shift due to the propagation from a point with radial coordinate u to one with $u+\Delta u$ reads ($c = G = 1$): According to (\ref{YR_43}), the proper time of observer $i$ will be related to the coordinate time $t$ by

\begin{eqnarray}
\frac{d\tau_i}{dt} = exp\Bigg(-\frac{\mu}{u}\Bigg).
\label{YR_4_10}
\end{eqnarray}

Suppose that the observer $\mathcal{O}_1$ emits a light pulse which travels to the observer $\mathcal{O}_2$, such that $\mathcal{O}_1$ measures the time between two successive crests of the light wave to be $\Delta \tau_1$. Each crest follows the same path to $\mathcal{O}_2$, except that they are separated by a coordinate time

\begin{eqnarray}
\Delta t = exp\Bigg(\frac{\mu}{u_1}\Bigg)\Delta \tau_1.
\label{YR_4_11}
\end{eqnarray}

This separation in coordinate time does not change along the photon trajectories, but the second observer measures a time between successive crests given by

\begin{eqnarray}
\Delta \tau_2 = exp\Bigg( - \frac{\mu}{u_2}\Bigg)\Delta t = \frac{ exp\Bigg(- \frac{\mu}{u_2}\Bigg)}{ exp\Bigg( - \frac{\mu}{u_1}\Bigg)} \Delta \tau_1.
\label{YR_4_12}
\end{eqnarray}

Since these intervals $\Delta \tau_i$ measure the proper time between two crests of an electromagnetic wave, the observed frequencies will be related by

\begin{eqnarray}
\frac{\omega_2}{\omega_1} = \frac{\Delta \tau_1}{\Delta \tau_2} =  \frac{ exp\Bigg(- \frac{\mu}{u_1}\Bigg)}{ exp\Bigg( - \frac{\mu}{u_2}\Bigg)}.
\label{YR_4_13}
\end{eqnarray}

Due to the spherical symmetry of the system, one can consider the equatorial plane, i.e., $\theta = \frac{\pi}{2}$ . The Lagrangian describing the geodesics in the spacetime (\ref{YR_43}) is given by

\begin{eqnarray}
2\mathcal{L} =- e^{-\frac{2\mu}{u}}\Bigg(\frac{dt}{d\lambda}\Bigg)^2 + e^{\frac{2\mu}{u}}\Bigg(\frac{du}{d\lambda}\Bigg)^2 + e^{\frac{2\mu}{u}}u^2\Bigg(\frac{d\varphi}{d\lambda}\Bigg)^2 = - \epsilon
\label{YR_4_14} 
\end{eqnarray}

where $\epsilon = -1, 0, +1$ for spacelike, null and timlike geodesics, respectively. To analyze ISCO, we take $\epsilon = 1$. In the metric (\ref{YR_43}) the lapse functions are independent of both $t$ and $\varphi$, so, there are two conserved quantities along the geodesic, i.e. energy $E$ and angular momentum $L$. From the Lagrangian (\ref{YR_4_14}), one can derive the generalized momenta

\begin{eqnarray}
p_t = - e^{-\frac{2\mu}{u}}\Bigg(\frac{dt}{d\lambda}\Bigg) = - E, \nonumber \\
p_u = e^{\frac{2\mu}{u}}\Bigg(\frac{du}{d\lambda}\Bigg), \nonumber \\
p_\varphi = e^{\frac{2\mu}{u}}u^2\Bigg(\frac{d\varphi}{d\lambda}\Bigg) = L 
\label{YR_4_15} 
\end{eqnarray}

Substituting Eq. (\ref{YR_4_15}) into Eq.(\ref{YR_4_14}), one finds

\begin{eqnarray}
- E^2 + \Bigg(\frac{du}{d\lambda}\Bigg)^2 + u^2L = - \epsilon e^{-\frac{2\mu}{u}}.
\label{YR_4_16} 
\end{eqnarray}
It looks even nicer if we rewrite it as

\begin{eqnarray}
\frac{1}{2}\Bigg(\frac{du}{d\lambda}\Bigg)^2 + V\left(u\right)_{eff} = \frac{1}{2}E^2, 
\label{YR_4_17} 
\end{eqnarray}

where

\begin{eqnarray}
V\left(u\right)_{eff} = \frac{e^{-\frac{2\mu}{u}}}{2}\Bigg(\frac{L^2}{e^{\frac{2\mu}{u}}u^2} + \epsilon\Bigg)
\label{YR_4_18} 
\end{eqnarray}

There are different curves $V\left(u\right)_{eff}$ for different values of L; for any one of these curves, the behavior of the orbit can be judged by comparing the $\frac{1}{2}E^2$ to $V\left(u\right)_{eff}$. The general behavior of the particle will be to move in the potential until it reaches a “turning point” where $V\left(u\right)_{eff} = \frac{1}{2}E^2$, where it will begin moving in the other direction. Sometimes there may be no turning point to hit, in which case the particle just keeps going. In other cases the particle may simply move in a circular orbit at radius $u_c = const$; this can happen if the potential is flat, $\frac{dV\left(u\right)_{eff}}{du} =0$.
Differentiating (\ref{YR_4_18}), we find that the circular orbits occur when

\begin{eqnarray}
\frac{\epsilon\mu u^2 e^{\frac{2\mu}{u}} - L^2\left(u - 2\mu\right)}{e^{\frac{4\mu}{u}}u^2} = 0
\label{YR_4_19} 
\end{eqnarray}

For $\epsilon = 0$ (for massless particle as photon), we can easily solve (\ref{YR_4_19}) to obtain

\begin{eqnarray}
u_{ph} = 2\mu
\label{YR_4_20} 
\end{eqnarray}

This shows a maximum of $V\left(u\right)_{eff}$ at $u = 2\mu$ for every $L$. This means that a photon can orbit forever in a circle at this radius, but any perturbation will cause it to fly away either to $u = 0$ or $u = \infty$.

When $\epsilon = 1$ (for massive particles), there is no simple analytic way of determining $u_c\left(\mu, L\right)$ as a function of $\mu$ and L.  We asuume a circular orbit at $u = u_c$, one can solve for the required angular momentum $L_c\left(\mu, u_c\right)$ as a function of $u_c$ and $\mu$. From (\ref{YR_4_19}) we get

\begin{eqnarray}
L_c = \pm \frac{u_ce^{\frac{\mu}{u_c}}}{\sqrt{u_c - 2\mu}}
\label{YR_4_21} 
\end{eqnarray}

Solving equation

\begin{eqnarray}
\frac{\partial L_c}{\partial u_c} = \pm\frac{\mu\left(8\mu^3 - 16\mu^2u_c +8\mu u_c^2 - u_c^3\right)e^{\frac{\mu}{u_c}}}{2\sqrt{\mu\left(u_c - 2\mu\right)}\left(u_c - 2\mu\right)^2u_c} = 0,
\label{YR_4_22} 
\end{eqnarray}

we obtain

\begin{eqnarray}
u_{ISCO} = \left(3 \pm \sqrt{5}\right)\mu
\label{YR_4_23} 
\end{eqnarray}

 Only the positive root is relevant (the negative root lies below $u_c = 2\mu$ where there are no circular orbits, stable or unstable). Consequently we identify the location for the massive particle ISCO as
\begin{eqnarray}
u_{ISCO} = \left(3 + \sqrt{5}\right)\mu
\label{YR_4_23B} 
\end{eqnarray}

\subsection{Stress-energy tensor and energy condutions}

Let us examine the Einstein field equations for Yilmaz-Rosen spacetime. We put (without loss of generality) $G=c=1$.  Mixed components of the stress-energy tensor are: $\rho = - T_t^t$, $p_\parallel = T_u^u$ and $p_\perp = T_\theta^\theta = T_\varphi^\varphi$. Using the mixed components $G_\mu^\nu = 8\pi T_\mu^\nu$, this yields the following form of the stress-energy momentum tensor:

\begin{eqnarray}
\rho = - \frac{\mu^2}{8\pi u^4}e^{-\frac{2\mu}{u}}, \quad p_\parallel =  - \frac{\mu^2}{8\pi u^4}e^{-\frac{2\mu}{u}} \quad p_\perp =   \frac{\mu^2}{8\pi u^4}e^{-\frac{2\mu}{u}} 
\label{YR_5_1R}
\end{eqnarray}

The energy condutions are just inequalities on $T_\mu^\nu$:

\textbf{1. Weak Energy condution (WEC):} $\rho \geq 0$, $\rho + p_\parallel \geq 0$ and $\rho + p_\perp \geq 0$;

\textbf{2. Null Energy condution (NEC):} $\rho + p_\parallel \geq 0$, $\rho + p_\perp \geq 0$;

\textbf{3. Strong Energy condution (SEC):} $\rho + p_\parallel  + 2p_\perp \geq 0$, $\rho + p_\parallel \geq 0$  and $\rho + p_\perp \geq 0$;

\textbf{4. Dominant Energy condution (DEC):} $\rho \geq p_\parallel$, $\rho \geq p_\perp$.

We see that, all the energy conditions are violated in the Yilmaz-Rosen metric. Such energy-momentum tensor corresponds to exotic matter with negative pressure (since $T_u^u$ is negative). This may be due to a scalar field (such as the Higgs field or another nonlinear field) or to "dark energy" or quintessence in some models.

\section{Yilmaz-Rosen solution  in the scalar-Einstein-Gauss-Bonnet (sEGB) $4d$ gravitational model}

As it is shown in our reconstruction in the scalar-Einstein-Gauss-Bonnet (sEGB) 4d model, unique solutions were found in the Yilmaz-Rosen metric. In the metric (\ref{2A}), the lapse functions can be selected as follows:

\begin{eqnarray}
A\left(u\right) = e^{2\alpha\left(u\right)},  \nonumber \\  C\left(u\right) = e^{-2\alpha\left(u\right)}\cdot u^2, \\ \alpha\left(u\right) = -\frac{\mu}{u}, \nonumber
\label{YR5_A} 
\end{eqnarray}

where $\mu >0$. From there it is clear that for $u \to +\infty$  the lapse functions of the metric $A\left(u\right) \to 1$, $C\left(u\right) \sim u^2$, i.e., in these domains the metric is asymptotically flat. Expanding the lapse functions $A\left(u\right)$  and $C\left(u\right)$ into asymptotic series, we see that for big enough $u$:

\begin{eqnarray}
A\left(u\right) = 1 - \frac{2\mu}{u}  + O\big(u^{-2}\big), \nonumber \\ 
C\left(u\right) \sim u^2. \label{YR21A}
\end{eqnarray}

In Yilmaz-Rosen metric for the "master equations" (\ref{6FF}) of the functions $E(u)$, $F(u)$ and $G(u)$, which defined in (\ref{6E}), (\ref{6F6}) and (\ref{6G6}), are respectively expressed as

\begin{equation}
E = -\frac{48\mu}{u^2}\bigg(u - \frac{2\mu}{3}\bigg)e^{2\alpha\left(u\right)},  
\label{EL_Bron_4} 
\end{equation}

\begin{equation}
F =  \frac{64\mu}{u^4}\bigg(u - 2\mu \bigg)\bigg(u - \frac{\mu}{2}\bigg)e^{2\alpha\left(u\right)},  
\label{EL_Bron_5} 
\end{equation}

\begin{equation}
G = 0.  
\label{EL_Bron_6} 
\end{equation}

Solving master equation $ E\ddot{f} + F\dot{f} + G = 0$ we obtain

\begin{equation}
f\left( u \right) =C_0 \int\frac{u^2}{\left(3u - 2\mu\right)^{\frac{2}{3}}} e^{-2\alpha\left(u\right)}du + C_1
\label{EL_Bron_7} 
\end{equation}

and

\begin{eqnarray}
\dot{f} = \frac{C_0u^2}{\left(3u - 2\mu\right)^{\frac{2}{3}}} e^{-2\alpha\left(u\right)},  \qquad
\ddot{f} =\frac{4C_0\bigg(2u - \mu \bigg)\bigg(\frac{u}{2} - \mu\bigg)}{\left(3u - 2\mu\right)^{\frac{5}{3}}} e^{-2\alpha\left(u\right)}
\label{EL_Bron_8A}
\end{eqnarray}

where $C_0$ and  $C_1$ are arbitrary constants.

For the potential $U\left(u\right)$  in the expressions (\ref{6EU}), (\ref{6FU}), (\ref{6GU}) the following expressions are valid:

\begin{eqnarray}
E_U = -\Bigg(\frac{2}{u}\Bigg)^2\left(2u - \mu\right)\mu e^{-2\alpha\left(u\right)}, \quad  \label{EL_Bron_9A} \\
F_U =-\frac{16\mu}{u^4}\Bigg(u^2 - \frac{7\mu}{2}u +2\mu^2\Bigg) ,  \quad \label{EL_Bron_10A} \\
G_U = 0  \qquad  \label{EL_Bron_11A} 
\end{eqnarray}
and hence using (\ref{6UU}) and (\ref{EL_Bron_8A}) we get the following relation for potential function

\begin{eqnarray}
U =16C_0M\frac{\left(u - \mu\right)\left(u - \frac{\mu}{2}\right)\left(u - 2\mu\right)}{u^4\left(3u - 2\mu\right)^{\frac{5}{3}}}e^{2\alpha\left(u\right)} \quad 
\label{EL_Bron_12A}
\end{eqnarray}

As $u \to + \infty$, the value of the potential function tends to zero. 

The relation (\ref{6P2}) reads in this case as follows

\begin{eqnarray}
\varepsilon \dot{\varphi}^2 = h\left(u\right),
\label{EL_Bron_13A}
\end{eqnarray}

\begin{eqnarray}
h\left(u\right) =\frac{\mu}{u^4}\Bigg( \frac{16C_0\mu\left(4\mu^3 - 14\mu^2u + 15\mu u^2 - 4\mu^3\right)}{\left(3u - 2\mu\right)^{\frac{5}{3}}}- 2\mu\Bigg). \quad  
\label{EB12BC}
\end{eqnarray}

For the general case, we assume $C_0 \neq 0$. From formula (\ref{EB12BC}) we obtain asymptotic relations as $u \to + \infty$

\begin{eqnarray}
h\left(u\right) \sim - \Bigg(\frac{64C_0\mu^{\frac{1}{3}}}{3^{\frac{5}{3}}}\Bigg)\cdot u^{-\frac{8}{3}}.
\label{EL_Bron_15A}
\end{eqnarray}

From (\ref{EL_Bron_15A}) we see, that for big enough $ \left|u\right|$ the

\begin{eqnarray}
sign\left(h\left(u\right)\right) = - sign\left(C_0\right),
\label{EL_Bron_16}
\end{eqnarray}
which can be written as
\begin{eqnarray}
h\left(u\right) < 0
\label{EL_Bron_17A}
\end{eqnarray}
for $C_0 >0$ and
\begin{eqnarray}
h\left(u\right) > 0
\label{EL_Bron_18A}
\end{eqnarray}
for $C_0 <0$.

This means that in the case of $C_0 > 0$, as it is shown in (\ref{EL_Bron_17A}), we obtain a ghost field ($\varepsilon = -1$) at big enough $u$, and in the case of $C_0 <0$ , as it is shown in (\ref{EL_Bron_18A}), we obtain an ordinary scalar field for big enough $u$.

At the point

\begin{eqnarray}
u_{\ast} = \frac{2\mu}{3}
\label{EL_Bron_19A}
\end{eqnarray}
the function $h\left(u\right)$ is not formally defined. We calculate the value of the numerator in the second term in brackets on the right-hand side of relation (\ref{EB12BC}) at the point $u = u_{\ast}$.  In formula (\ref{EB12BC}) and everywhere in this article we assume

\begin{eqnarray}
x^{\frac{5}{3}} \equiv \left(\sqrt[3]x\right)^5
\label{EL_Bron_20A}
\end{eqnarray}
for any real number $x$. We denote
\begin{eqnarray}
N\left(u\right) = 16C_0\mu\left(4\mu^3 - 14\mu^2u + 15\mu u^2 - 4\mu^3\right),
\label{EL_Bron_21A}
\end{eqnarray}
and find value of the function $N\left(u\right)$ at the point $u_{\ast}$: 

\begin{eqnarray}
N\left(u_{\ast}\right) = \frac{64C_0M^3}{27}.
\label{EL_Bron_22A}
\end{eqnarray}

From this formula follows the following asymptotic relationship for $u \to u_{\ast}$:

\begin{eqnarray}
h\left(u\right) \sim  \frac{1}{u_{\ast}^4} \cdot \frac{64C_0M^3}{27} \cdot \frac{1}{\left(u_{\ast} - u\right)^{\frac{5}{3}}}
\label{EL_Bron_23A}
\end{eqnarray}

By definition (\ref{EL_Bron_20A}), the function $\left(u_{\ast} - u\right)^{-\frac{5}{3}} $ tends to $+\infty$  as $u \to u_{\ast} - 0$  (on the left) and to  $-\infty$ as $u \to u_{\ast} + 0$  (on the right). Therefore, from relation (\ref{EL_Bron_23A}) we obtain

\begin{eqnarray}
h\left(u\right) \to \left(-sign\left(C_0\right)\right)\infty
\label{EL_Bron_24A}
\end{eqnarray}
as $u \to u_{\ast} - 0$  and
\begin{eqnarray}
h\left(u\right) \to \left(sign\left(C_0\right)\right)\infty
\label{EL_Bron_25A}
\end{eqnarray}
as $u \to u_{\ast} + 0$.

Thus $u_{\ast}$ is a breaking point of the function $h\left(u\right)$. From these considerations the following statement is followed: There is no such reconstruction constant $C_0 \neq 0$ for the Yilmaz-Rosen metric with parameter $\mu > 0$  that for all values  $u \in \left( 0,+\infty\right)$ except the point $u_{\ast}$ the function $h\left(u\right)$ is of constant sign, i.e. $h\left(u\right) >0$ $\left(\varepsilon = 1 \right)$ or $h\left(u\right) <0$ $\left(\varepsilon =-1 \right)$.

From a physical point of view, this means that for any non-trivial reconstruction ($C_0 \neq 0$) \cite{Er_Ivash} of the Yilmaz-Rosen metric with parameter $\mu > 0$, the scalar field defined in the interval  $u \in \left( 0,+\infty\right)$ cannot be purely ghost or purely normal (non-ghost). 

In the case of a trivial reconstruction of the Yilmaz-Rosen metric (with $\mu > 0$) with the reconstruction constant $C_0 = 0$  the Gauss-Bonnet term gives a zero contribution to the equations of motion and we arrive a solution in the Yilmaz-Rosen metric with a purely ghost scalar field ($\varepsilon = - 1$) and zero potential $U = 0$.

Figures 1 and 2 show examples for the function $h\left(u\right)$ with  $C_0 < 0$ and  $C_0 > 0$, respectively.

\begin{figure}[h]
\center{\includegraphics[scale=0.15]{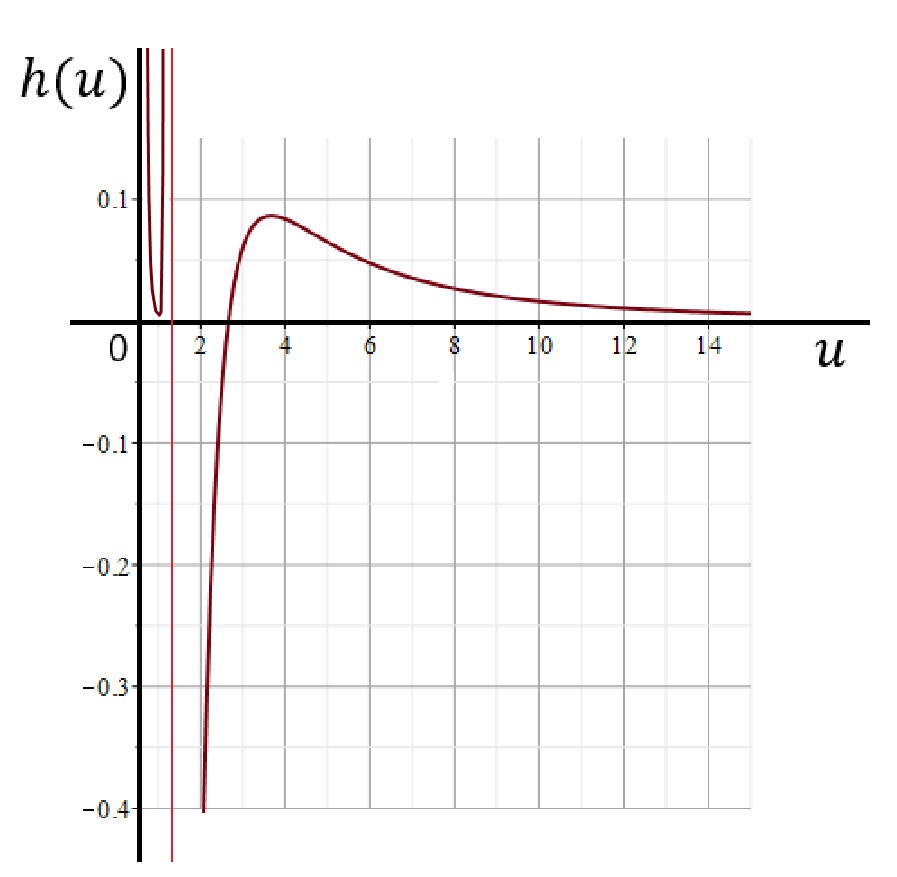}}
\caption{\textsl{The function $h\left(u\right)$   for $\mu = 1$ and $C_0 = -10$. A vertical red dashed line crosses point $u_{\ast}=\frac{1}{3}$. A) The function $h\left(u\right)$ is positive in the interval $\left(0, u_{\ast}=\frac{2}{3}\right)$. This means that in this interval a scalar field is ordinary one. B) The function $h\left(u\right)$ is negative in the interval  $\left(u_{\ast}=\frac{1}{3}, u_1=3.4686\right)$ and positive in the interval $\left(u_1=3.4686, +\infty\right)$. Therefore in the interval $\left(u_{\ast}=\frac{1}{3}, u_1=3.4686\right)$ a scalar field is ghost one and in the interval $\left(u_1=3.4686, +\infty\right)$ we obtain a solution with an ordinary field.}} 
\label{Fig_1}
\end{figure}

\begin{figure}[h]
\center{\includegraphics[scale=0.15]{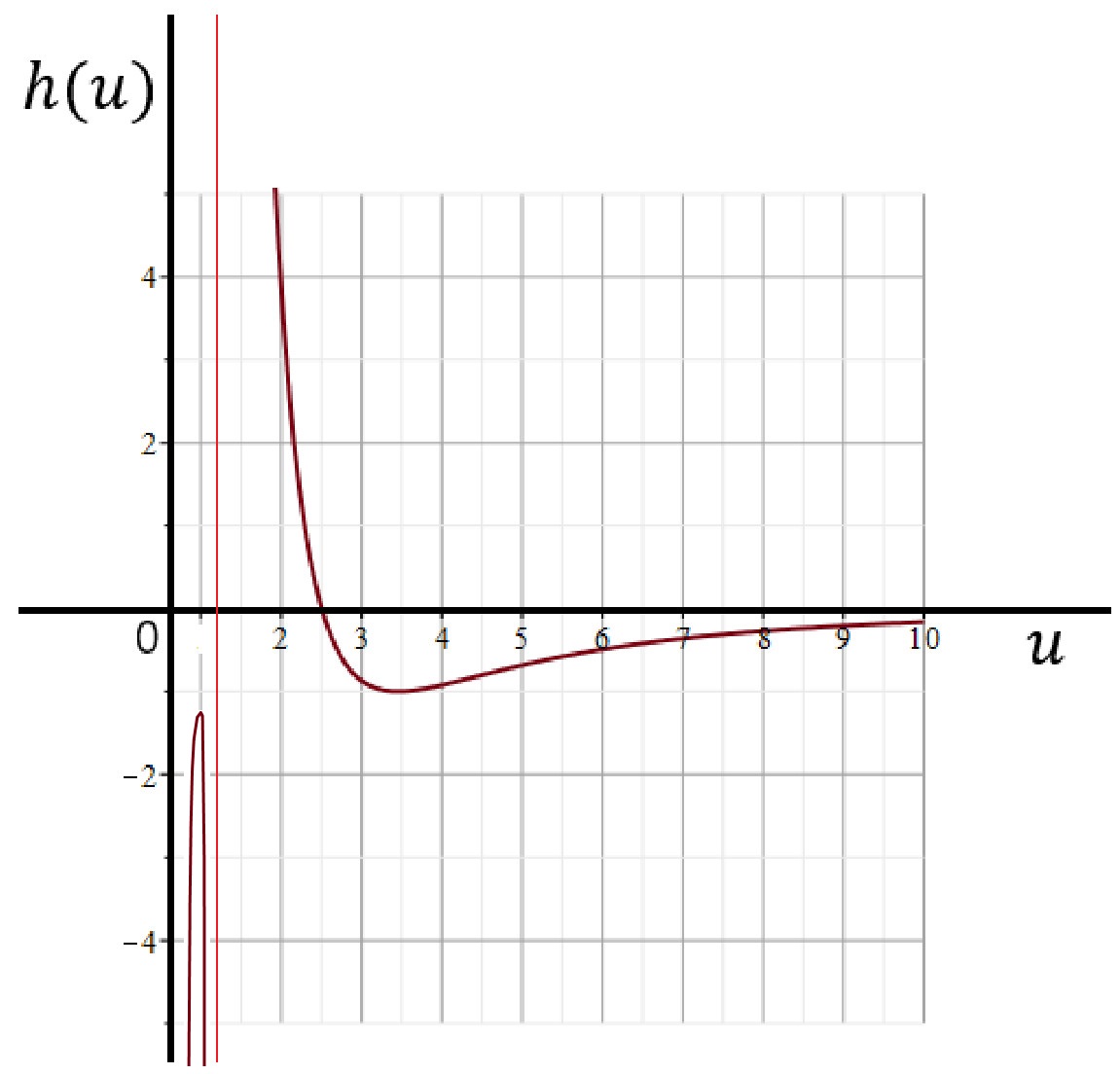}}
\caption{\textsl{The function $h\left(u\right)$  for $\mu = 1$ and $C_0 = 10$.  A vertical red dashed line crosses point $u_{\ast}=\frac{1}{3}$. A) The function $h\left(u\right) < 0$ in the interval $\left(0, u_{\ast}=\frac{1}{3}\right)$. This means that in this interval a scalar field is ghost one. B) The function $h\left(u\right)$ is positive in the interval  $\left(u_{\ast}=\frac{1}{3}, u_1=2.1496\right)$ and negative in the interval  $\left(u_1=2.1496, +\infty\right)$. Therefore only in the interval $\left(u_{\ast}=\frac{1}{3}, u_1=2.1496\right)$ we obtain a solution with an ordinary field.}} 
\label{Fig_2}
\end{figure}

From Fig.1 we see that, for $\mu = 1$ and $C_0 = - 10$ in the interval  $u \in \left( u_{\ast} = \frac{2}{3}, u_1 = 2.5324 \right)$ we obtain a solution with a ghost field, and in the intervals $u \in \left(0, u_{\ast} = \frac{2}{3}\right)$ and $u \in \left( u_1 = 2.5324, + \infty \right)$ we obtain a solution with an ordinary field.

From Fig.2 we see that, for $\mu = 1$ and $C_0 = 10$ in the interval  $u \in \left( u_{\ast} = \frac{2}{3}, u_1 = 2.5324 \right)$ we obtain a solution with an ordinary field, and in the intervals $u \in \left(0, u_{\ast} = \frac{2}{3}\right)$ and $u \in \left( u_1 = 2.5324, + \infty \right)$ we obtain a solution with a ghost field.

\subsection{The Yilmaz-Rosen solution  in the Scalar-Tensor theory with minimal coupling}

On the base of the scalar-tensor theory, we can find a solution in the Yilmaz-Rosen metric. In this case, the potential is determined by formula (\ref{ST_2A}) and is equal to zero

\begin{eqnarray}
U = 0
\label{ST_1AB}
\end{eqnarray}

From relation (\ref{ST_3A}) we get

\begin{eqnarray}
\varepsilon\dot{\varphi}^2 = h\left(u\right) = -\frac{2\mu^2}{u^4}
\label{ST_2AB}
\end{eqnarray}
which urges the scalar field to be phantom one $(\varepsilon = -1)$, and

\begin{eqnarray}
\dot{\varphi} = \pm \frac{\sqrt{2}\mu}{u^2}
\label{ST_3AB}
\end{eqnarray}

or, equivalently

\begin{eqnarray}
\varphi - \varphi_0 = \pm \frac{\sqrt{2}\mu}{u}
\label{ST_4AB}
\end{eqnarray}

\section{The Janis–Newman–Winicour (JNW) spacetime}

The Janis–Newman–Winicour (JNW) metric \cite{JNW_1} is a static, spherically symmetric solution to the Einstein field equations coupled to a massless scalar field. It is perhaps the most famous and simplest generalization of the Schwarzschild metric that includes a scalar "hair." In standard general relativity (GR), the no-hair theorem states that stationary black holes in vacuum are completely characterized by only three parameters: mass, electric charge, and angular momentum (summarized by the phrase "black holes have no hair"). The Schwarzschild metric describes the simplest case: a neutral, non-rotating black hole defined solely by its mass $M$.

The JNW solution arises when you challenge the vacuum assumption by introducing a massless scalar field $\varphi$. This provides a simple model to explore how matter fields can endow a black hole-like object with additional structure or "hair." The JNW metric is derived by solving the Einstein equations where the energy-momentum tensor $T_{\mu\nu}$ is generated by a massless, minimally coupled scalar field $\varphi$. The JNW metric is given by

\begin{eqnarray}
ds^2 =-\Bigg(1 - \frac{b}{u}\Bigg)^sdt^2  + \Bigg(1 - \frac{b}{u}\Bigg)^{-s}du^2 + u^2\Bigg(1 - \frac{b}{u}\Bigg)^{1 - s}d\Omega^2  \label{JNW_1} 
\end{eqnarray}

By expanding the metric in the order of $b/r$, we find that the parameter b is related to the physical mass of the spacetime $M$ by

\begin{eqnarray}
b = \frac{2\mu}{s} \nonumber
\end{eqnarray}

When $s = 1$, it is exactly the Schwarzschild radius. Then, the JNW solution becomes the following expression:

\begin{eqnarray}
ds^2 =-\Bigg(1 - \frac{2\mu}{s\cdot u}\Bigg)^sdt^2  + \Bigg(1 - \frac{2\mu}{s \cdot u}\Bigg)^{-s}du^2 + u^2\Bigg(1 - \frac{2\mu}{s\cdot u}\Bigg)^{1 - s}d\Omega^2  \label{JNW_2} 
\end{eqnarray}

The most striking feature of the JNW spacetime, and the core of its physical significance, is the nature of its central singularity. In the Schwarzschild metric ($s=1$), the surface $u = b = 2M$ is a coordinate singularity - the event horizon -which cloaks the unavoidable curvature singularity at $u = 0$.

In case of $s < 1$, the surface $u = b$ is no longer a horizon but a naked singularity. The curvature invariant $R^{\mu\nu\rho\sigma} R_{\mu\nu\rho\sigma}$ (the Kretschmann scalar) diverges at both $u = b$ and $u = 0$. Crucially, the surface $u = b$ is a point-like singularity in all directions except one, making it a timelike naked singularity. This means that it is not hidden from external observers and that its effects can, in principle, be observed and interacted \cite{Virb}. The absence of an event horizon is a direct consequence of the scalar field's stress-energy, which violates the energy conditions necessary for the black hole no-hair theorems to hold. The JNW metric thus serves as a robust counterexample to the cosmic censorship hypothesis, which conjectures that all singularities are hidden behind horizons.

The JNW metric is not a black hole metric (except in the trivial limit where it becomes one). It is the metric for a naked singularity surrounded by a scalar field. Its primary importance lies in its role as a simple, exact solution that illustrates the consequences of introducing new fundamental fields into GR and provides a theoretical laboratory for studying naked singularities and violations of cosmic censorship.

Using a calculation method analogous to the Yılmaz-Rosen metric, we find that the photon sphere radius and ISCO for the JNW metric are:
\begin{equation}
u_{ph} = \frac{(2s + 1)\mu}{s}  
\label{JNW_2AA} 
\end{equation}
and
\begin{equation}
u_{ISCO} = \Bigg(3+ \frac{1}{s} + \sqrt{5 - \frac{1}{s^2}}\Bigg)\mu  
\label{JNW_2BB} 
\end{equation}

It is impotant to point out that, Yilmaz-Rosen metric is obtained by putting $s \to +\infty$ in the JNW metric (\ref{JNW_2}):

\begin{eqnarray}
\lim_{s\to +\infty} \Bigg(1 - \frac{2\mu}{su}\Bigg)^s = e^{-\frac{2\mu}{u}}, \nonumber \\
\lim_{s\to +\infty} \Bigg(1 - \frac{2\mu}{su}\Bigg)^{1-s}u^2 = e^{\frac{2\mu}{u}}u^2 \nonumber
\end{eqnarray}

In the same way, by putting $s \to +\infty$ in Equation (\ref{ST_JNW_2AB}), we obtain (\ref{ST_4AB}).

We have checked that the above metric and the phantom field do solve the Einstein equations and the scalar field equation. The metric is known as Yilmaz “exponential metric” in the literature [45–48]. But some researchers pointed out that it appeared for the first time in Ref. \cite{Papapetr} by Papapetrou. The metric has attracted wide attention in the literature \cite{Misner} - \cite{Robertson_2} ever since it was found. Very recently, Boonserm et al. \cite{Boon} showed that the Yilmaz-Rosen exponential metric represents a traversable wormhole. Traversable wormholes have been studied in many aspects, such as the stability analysis of wormholes \cite{Boon} - \cite{Li}, the resolution to the horizon problem in cosmology \cite{Hoch} - \cite{Kim}, the wormhole solutions in the modified gravities \cite{Jusuf} - \cite{Blaz} etc.

\section{The Janis-Newman-Winicour (JNW) solution  in the scalar-Einstein-Gauss-Bonnet (sEGB) $4d$ gravitational model}

In JNW metric for the "master equations" (\ref{6FF}) of the functions E(u), F(u) and G(u), which defined in (\ref{6E}), (\ref{6F6}) and (\ref{6G6}), are respectively expressed as:

\begin{equation}
E = -\frac{16\mu\Bigg(\left(3u - 2\mu\right)s^2 - \left(3s + 1\right)\mu\Bigg)\Bigg(1 - \frac{2\mu}{su}\Bigg)^s}{su\left(2\mu - su\right)},  
\label{JNW_3} 
\end{equation}

\begin{equation}
F =  \frac{32\mu\left(2s\mu - su + \mu\right)\left(\mu s^2 - 2s^2u + 2\mu s +\mu\right)}{\left(2\mu - su\right)^2u^2s},  
\label{JNW_4} 
\end{equation}

\begin{equation}
G = 0.  
\label{JNW_5} 
\end{equation}

The solution of the master equation (\ref{6FF}) is

\begin{equation}
f\left( u \right) =C_0 \int\frac{u^{1+s}\left(2\mu - su\right)^{1-s}}{\left(2\mu s^2 - 3s^2u + 3\mu s + \mu\right)^{\frac{2}{3}}} e^{-2\alpha\left(u\right)}du + C_1
\label{JNW_6} 
\end{equation}

and

\begin{eqnarray}
\dot{f} = \frac{C_0u^{1+s}\left(2\mu - su\right)^{1-s}}{\left(2\mu s^2 - 3s^2u + 3\mu s + \mu\right)^{\frac{2}{3}}}, \nonumber  \\
\ddot{f} =\frac{4C_0u^{1+s}\left(2\mu - su\right)^{1-s}\Bigg(\left(\mu - \frac{u}{2}\right)s + \frac{\mu}{2}\Bigg)\Bigg(\left(\mu - 2u\right)s^2 + 2\mu s +\mu\Bigg)}{\left(2\mu u - u^2s\right)\left(2\mu s^2 - 3s^2u + 3\mu s + \mu\right)^{\frac{5}{3}}} ,
\label{JNW_7}
\end{eqnarray}

where $C_0$ and  $C_1$ are arbitrary constants.

For the potential $U\left(u\right)$ in the expressions (\ref{6EU}), (\ref{6FU}), (\ref{6GU}) the following expressions are valid:

\begin{equation}
E_U = -\frac{4\mu\Bigg(\left(1 + s\right)^2\mu - 2s^2u\Bigg)\Bigg(1 - \frac{2\mu}{su}\Bigg)^s}{su\left(2\mu - su\right)},  
\label{JNW_8} 
\end{equation}

\begin{equation}
F_U =  \frac{8\mu\left(2s\mu - su + \mu\right)\left(\mu s^2 - 2s^2u + 2\mu s +\mu\right)\Bigg(1 - \frac{2\mu}{su}\Bigg)^s}{\left(2\mu - su\right)^2u^2s},  
\label{JNW_9} 
\end{equation}

\begin{equation}
G_U = 0.  
\label{JNW_10} 
\end{equation}

\subsection{s = 2}

In this case, a new metric describing the BBMB-like black hole is obtained from the JNW metric (\ref{JNW_2}). The BBMB black hole is an exact, static, and spherically symmetric solution to the equations of Einstein's theory of gravity when it is coupled to a special type of massless scalar field and an electromagnetic field. Its most revolutionary feature is that it is a black hole with "scalar hair”. The BBMB solution has the same mass and charge as a Reissner-Nordström black hole, but it also has an additional, independent parameter: the configuration of the scalar field surrounding it. This scalar field is not hidden behind the horizon; it persists outside, making the black hole "hairy".

The BBMB solution, derived independently by Bocharova, Bronnikov, and Melnikov in 1970 \cite{Bocharova} and later rediscovered by John Bekenstein \cite{Beken} in 1974, is a black hole model within the framework of Einstein–Maxwell conformal scalar gravity. Its essence lies in the coupling of a scalar field - a fundamental force or particle represented by a single value at every point in space - to gravity in a specific, conformally invariant way. Conformal invariance is a symmetry property meaning the theory's equations remain unchanged under transformations that locally rescale distances, a property deeply connected to the structure of spacetime itself.

What makes the BBMB black hole so remarkable is its composition and its apparent defiance of a long-held principle: the "no-hair" theorem. This theorem posits that black holes are simple objects, characterized only by their mass, electric charge, and angular momentum; all other information about the matter that formed them is lost behind the event horizon, succinctly summarized as "black holes have no hair." The BBMB solution challenges this by sporting a well-defined, non-trivial scalar field outside its event horizon. It possesses "hair" in the form of this scalar field, making it a primary exhibit in discussions about violations and generalizations of the no-hair theorem.

The geometry of the BBMB black hole is both intriguing and strangely familiar. Its spacetime metric is identical to that of an extremal Reissner–Nordström black hole. An extremal black hole is one whose charge is equal to its mass (in natural units), and it represents a limiting case where the inner and outer horizons merge into a single horizon. This means the BBMB black hole has a single event horizon, and its gravitational field is indistinguishable from a maximally charged black hole. However, its internal structure is vastly different. The scalar field, which is smooth everywhere outside the horizon, becomes singular on the horizon itself. This is a paradoxical feature: the event horizon, a one-way membrane in spacetime, also serves as a surface of infinite energy density for the scalar field.

Let us consider the BBMB-like black hole metric solutions in the scalar-Einstein-Gauss-Bonnet (sEGB) $4d$ gravitational model. The BBMB-like metric is given by 

\begin{eqnarray}
ds^2 =-\Bigg(1 - \frac{\mu}{u}\Bigg)^2dt^2  + \Bigg(1 - \frac{\mu}{u}\Bigg)^{-2}du^2 + \frac{u^2}{\Bigg(1 - \frac{\mu}{u}\Bigg)}d\Omega^2  \label{JNW_11} 
\end{eqnarray}

In this case, the potential $U\left(u\right)$ in (\ref{6UU}) is equal to the following:

\begin{eqnarray}
U\left(u\right) = -\frac{2C_0\mu\Bigg(\frac{135}{2}\mu^5u^3 - 267\mu^4u^4 + \frac{827}{2}\mu^3u^5 - 312\mu^2u^6 + 114\mu u^7 - 16u^8\Bigg)}{u^7\left(u - \mu\right)^2\left(15\mu - 12u\right)^{\frac{5}{3}}} \qquad \label{JNW_12} 
\end{eqnarray}

The relation (\ref{6P2}) reads in this case as follows

\begin{eqnarray}
\varepsilon \dot{\varphi}^2 = h\left(u\right),
\label{JNW_13}
\end{eqnarray}

\begin{eqnarray}
h\left(u\right) = \frac{\mu\Bigg(540C_0\left(\mu^3 - 2\mu^2 u + \frac{34}{27}\mu u^2 - \frac{32}{135}u^3\right) + \mu\left(18u - \frac{45\mu}{2}\right) \left(15\mu - 12u\right)^{\frac{2}{3}}\Bigg)}{u^2\left(u - \mu\right)^2\left(15\mu - 12u\right)^{\frac{5}{3}}} \quad  
\label{JNW_14}
\end{eqnarray}

Let $C_0 \neq 0$. From this formula we obtain asymptotic relations as  $u \to +\infty$

\begin{eqnarray}
h\left(u\right) \sim \Bigg(\frac{16C_0\mu}{3\left(18\right)^{\frac{1}{3}}}\Bigg)u^{-\frac{8}{3}}
\label{JNW_14A}
\end{eqnarray}

From (\ref{JNW_14A}) we see, that for big enough $u$ the

\begin{eqnarray}
sign\left(h\left(u\right)\right) = sign\left(C_0\right),
\label{EL_Bron_16}
\end{eqnarray}
which can be written as
\begin{eqnarray}
h\left(u\right) < 0
\label{EL_Bron_17}
\end{eqnarray}
for $C_0 < 0$ and
\begin{eqnarray}
h\left(u\right) > 0
\label{EL_Bron_18}
\end{eqnarray}
for $C_0 >0$.

This means that in the case of $C_0 < 0$, due to (\ref{EL_Bron_17}), we obtain a ghost field $\left(\varepsilon = -1\right)$ for big enough $u$, and in the case of $C_0 > 0$, due to (\ref{EL_Bron_18}), we obtain an ordinary scalar field for big enough $u$.

At the points

\begin{eqnarray}
u_1 = 0, \nonumber \\
u_2 = \mu, \nonumber \\
u_{3} = \frac{5\mu}{4}
\label{EL_Bron_19}
\end{eqnarray}
the function $h\left(u\right)$ is not formally defined. We calculate the value of the numerator in (\ref{JNW_14}) at the points $u_1, u_2$ and $u_3$.  In formula (\ref{JNW_14}) and everywhere in this article we assume

\begin{eqnarray}
x^{\frac{5}{3}} \equiv \left(\sqrt[3]x\right)^5
\label{EL_Bron_20}
\end{eqnarray}
for any real number $x$. Let us denote
\begin{eqnarray}
N\left(u\right) = 540C_0\left(\mu^3 - 2\mu^2 u + \frac{34}{27}\mu u^2 - \frac{32}{135}u^3\right) + \mu\left(18u - \frac{45\mu}{2}\right) \left(15\mu - 12u\right)^{\frac{2}{3}} 
\label{EL_Bron_21}
\end{eqnarray}
and find the value of the function $N\left(u\right)$ at the points $u_1, u_2$ and $u_3$. We obtain 

\begin{eqnarray}
N\left(u_1\right) = 45\mu^4\Bigg(12C_0 - \frac{15^{\frac{2}{3}}}{2\mu^{\frac{1}{3}}}\Bigg), \nonumber \\
N\left(u_2\right) = 3\mu^4\Bigg(4C_0 - \frac{3^{\frac{5}{3}}}{2\mu^{\frac{1}{3}}}\Bigg), \nonumber \\
N\left(u_{3}\right) = \frac{5C_0\mu^4}{2}.
\label{EL_Bron_22}
\end{eqnarray}

From this formula follows the following asymptotic relationship for $u \to u_{1}$:

\begin{eqnarray}
h\left(u\right) \sim  \frac{\mu N\left(u_1\right)}{\left(u_1 - u\right)^2\left(u_1 - \mu\right)\left(15\mu - 12u_1\right)^{\frac{5}{3}}} =  \frac{ 45\mu^5\Bigg(12C_0 - \frac{15^{\frac{2}{3}}}{2\mu^{\frac{1}{3}}}\Bigg)}{\left(u_1 - u\right)^2\left(u_1 - \mu\right)\left(15\mu - 12u_1\right)^{\frac{5}{3}}} 
\label{EL_Bron_23}
\end{eqnarray}

The functions $\left(u_1 - u\right)^2 >0$, $\left(u_{1} - \mu\right) < 0$ and $\left(15\mu - 12u_1\right)^{\frac{5}{3}} >0$ Therefore, in case of $C_0 > \frac{15^{\frac{2}{3}}}{24\mu^{\frac{1}{3}}}$

\begin{eqnarray}
h\left(u\right) \to - \infty
\label{EL_Bron_24}
\end{eqnarray}
and
\begin{eqnarray}
h\left(u\right) \to + \infty
\label{EL_Bron_25}
\end{eqnarray}
when $C_0 < \frac{15^{\frac{2}{3}}}{24\mu^{\frac{1}{3}}}$.

We obtain  the following asymptotic relationship for $u \to u_{2}$:

\begin{eqnarray}
h\left(u\right) \sim  \frac{\mu N\left(u_2\right)}{u_2^2\left(u_2 - u\right)^2\left(15\mu - 12u_2\right)^{\frac{5}{3}}} = \frac{ \mu^{\frac{4}{3}}\Bigg(4C_0 - \frac{3^{\frac{5}{3}}}{2\mu^{\frac{1}{3}}}\Bigg)}{3^{\frac{2}{3}}\left(u_2 - u\right)^2} 
\label{EL_Bron_26}
\end{eqnarray}

It is seen from the last formula that, in case of $C_0 > \frac{3^{\frac{5}{3}}}{8\mu^{\frac{1}{3}}}$

\begin{eqnarray}
h\left(u\right) \to + \infty
\label{EL_Bron_27}
\end{eqnarray}

as $u \to u_2 - 0$ and $u \to u_2 + 0$.

\subsection{$s = \frac{1}{2}$}

In this case the JNW metric is given by

\begin{eqnarray}
ds^2 =-\Bigg(1 - \frac{4\mu}{u}\Bigg)^{\frac{1}{2}}dt^2  + \Bigg(1 - \frac{4\mu}{u}\Bigg)^{-\frac{1}{2}}du^2 + u^2\Bigg(1 - \frac{4\mu}{u}\Bigg)^{\frac{1}{2}}d\Omega^2,   \label{JNW_15} 
\end{eqnarray}

and the potential $U\left(u\right)$ in (\ref{6UU}) is equal to

\begin{eqnarray}
U\left(u\right) = \frac{C_0\mu\left(u - 3\mu\right)\left(2u -9\mu\right)}{256\left(8\mu - 2u\right)^{\frac{1}{2}}\left(3\mu - \frac{3u}{4}\right)^{\frac{5}{3}}u^{\frac{5}{2}}} \label{JNW_16} 
\end{eqnarray}

The function $h\left(u\right)$, which defined by (\ref{6P2}) is given by the following relation

\begin{eqnarray}
h\left(u\right) = \frac{\mu^2}{\left(4\mu - u\right)^2u^2}\Bigg(\frac{C_0\left(54\mu^2 - 27\mu u + 4u^2\right)\left(u - 4\mu\right)^{\frac{1}{2}}}{16\cdot 2^{\frac{1}{6}}\cdot 3^{\frac{5}{3}}\cdot \left(4\mu - u\right)^{\frac{7}{6}}} + 6\Bigg)
\label{JNW_16A}
\end{eqnarray}

From this formula we obtain asymptotic relations as  $u \to +\infty$

\begin{eqnarray}
h\left(u\right) \sim \Bigg(\frac{C_0\mu^2}{2^{\frac{13}{6}}\cdot 3^{\frac{5}{3}}}\Bigg)u^{-\frac{16}{6}}
\label{JNW_16B}
\end{eqnarray}

From (\ref{JNW_16B}) we see, that for big enough $ \left|u\right|$ the

\begin{eqnarray}
sign\left(h\left(u\right)\right) = sign\left(C_0\right),
\label{EL_Bron_16BC}
\end{eqnarray}
which can be written as
\begin{eqnarray}
h\left(u\right) < 0
\label{EL_Bron_17BC}
\end{eqnarray}
for $C_0 < 0$ and
\begin{eqnarray}
h\left(u\right) > 0
\label{EL_Bron_18BC}
\end{eqnarray}
for $C_0 >0$.

This means that in the case of $C_0 < 0$, due to (\ref{EL_Bron_17BC}), we obtain a ghost field $\left(\varepsilon = -1\right)$ for big enough $ u$, and in the case of $C_0 > 0$, due to (\ref{EL_Bron_18BC}), we obtain an ordinary scalar field for big enough $u$.

As it is seen from (\ref{JNW_16A}) in the interval $u \in \left(0, 4\mu\right)$  the function $h\left(u\right)$ is not formally defined. The naked singularity is occured at the point $4\mu$.

\subsection{The Janis-Newman-Winicour (JNW) solution in the Scalar-Tensor theory with minimal coupling}

In this subsection, we consider solution in the Scalar-Tensor theory by starting from the JNW spacetime. In this case, the potential is determined by formula (\ref{ST_2A}) and is equal to zero

\begin{eqnarray}
U = 0
\label{ST_JNW_1AB}
\end{eqnarray}

From relation (\ref{ST_3A}) we get

\begin{eqnarray}
\varepsilon\dot{\varphi}^2 = 2\left(1 - s^2\right)\Bigg(\frac{\mu}{\left(2\mu - su\right)u}\Bigg)^2
\label{ST_JNW_2AB}
\end{eqnarray}

In case of an ordinary scalar field $\varepsilon = 1$, and by integrating (\ref{ST_JNW_2AB}), we get

\begin{eqnarray}
\varphi = -\sqrt{\frac{\left(1 - s^2\right)}{2}}ln\Bigg(\frac{2\mu}{su} - 1\Bigg)
\label{ST_JNW_3A}
\end{eqnarray}

As it is shown from the last formula $s$ must be $s < 1$. For example, in case of $s = \frac{1}{2}$ a scalar field is defined as an ordinary one and equal to

\begin{eqnarray}
\varphi = -\frac{1}{2}\sqrt{\frac{3}{2}}ln\Bigg(\frac{4\mu}{u} - 1\Bigg)
\label{ST_JNW_3AB}
\end{eqnarray}

In case of a phantom scalar field $\varepsilon = -1$, and by integrating (\ref{ST_JNW_2AB}), we get

\begin{eqnarray}
\varphi = -\sqrt{\frac{\left(s^2 - 1\right)}{2}}ln\Bigg(\frac{2\mu}{su} - 1\Bigg)
\label{ST_JNW_4AB}
\end{eqnarray}

As it is shown from the last formula $s$ must be $s > 1$. For example, in BBMB-like black hole solution $s = 2$, and scalar field is phantom one and equal to

\begin{eqnarray}
\varphi = -\sqrt{\frac{3}{2}}ln\Bigg(\frac{\mu}{u} - 1\Bigg)
\label{ST_JNW_5AB}
\end{eqnarray}

\section{Conclusions}

The Yilmaz-Rosen metric represents a serious theoretical attempt to address perceived shortcomings in the standard formulation of general relativity. While empirically less successful than Einstein's theory to date, it offers valuable insights into gravitational energy, singularity structure, and the relationship between geometry and physics. The metric's mathematical elegance and distinct physical predictions ensure its continued study as part of the broader landscape of gravitational theories. Future observational advances may either further constrain or potentially revive interest in this alternative approach to gravitation. 

In this paper we consider the sEGB model - a $4d$ gravitational model with a scalar field $\varphi$, Einstein and Gauss-Bonnet terms. The model action contains a potential term $U\left(\varphi\right))$, a Gauss-Bonnet coupling function $f\left(\varphi\right)$ and a parameter $\varepsilon = \pm 1$, where $\varepsilon = 1$ corresponds to a pure scalar field, and $\varepsilon = -1$ to a phantom one.

Here we applied the sEGB reconstruction procedure considered in our previous paper to the Yilmaz-Rosen metric solution, which may describes a quasi black hole without event horizon. Along with it, analytical solutions based on the scalar-tensor theory in the Yılmaz-Rosen metric are obtained. As the obtained results show, in this case, the potential $U$ vanishes and the scalar field is being a phantom one.

Furthermore, the solution of Einstein equation in the Yılmaz-Rosen metric was considered. As it is shown in the obtained results, all energy conditions are violated. Such energy-momentum tensor may corresponds to exotic matter with negative pressure (since $T_u^u$ is negative). This may be due to a scalar field (such as the Higgs field or another nonlinear field) or to "dark energy" or quintessence in some models.

For this metric, written in the Buchdal parameterization with a radial variable $u$, we found a solution of the master equation for $f\left(\varphi\left(u\right)\right)$ with the integration (reconstruction) parameter $C_0$. Also, expressions for $U\left(\varphi\left(u\right)\right)$ and $\varepsilon \dot{\varphi} = h\left(u\right)$ were found for $\varepsilon = \pm 1$.
We found that for big enough values of  $u$: the scalar field is an ordinary field in case a) $C_0 < 0$, while it is a phantom one when b) $C_0 > 0$. It should be noted that case a) is more acceptable from a physical point of view. In this case, the value of the field potential $U\left(\varphi\left(u\right)\right)$ tends to zero as $u \to +\infty$, independently from the sign of an arbitrary constant $C_0 \neq 0$. The obtained results in this metric reported that  there is does not exist as a constant-sign function $h\left(u\right)$ for all values of $u \in \left(0, + \infty \right)$.

The obtained results have proved that for all non-trivial values of the parameter $C_0 \neq 0$ the function $h\left(u \right)$ is not of constant sign for all admissible $u \in \left(0, + \infty \right)$ (no-go theorem). This means that for a fixed value of the parameter $\varepsilon = \pm 1$ there is no non-trivial sEGB reconstruction in which the scalar field is a purely ordinary field ($\varepsilon = 1$) or a purely phantom one ($\varepsilon = -1$).

Subsequently, we applied the reconstruction method to the Janis-Newman-Winicour (JNW), which yielded the most recent results. As shown in this article, the Yılmaz-Rosen metric can be considered as a limiting case of the JNW metric as $s \to +\infty$. As shown in the final results, for any non-trivial reconstruction ($C_0 \neq 0$) \cite{Er_Ivash} of the JNW metric with parameter $\mu > 0$, the scalar field defined in the interval  $u \in \left( 0,+\infty\right)$ cannot be purely ghost or purely normal (non-ghost).

\section*{Acknowledgements}

I thank prof. V.D. Ivashchuk for helpful discussions and very useful comments.     

\section{Data Availability Statements}

No Data associated in the manuscript.

\renewcommand{\theequation}{\Alph{subsection}.\arabic{equation}}
\renewcommand{\thesection}{}
\renewcommand{\thesubsection}{\Alph{subsection}}
\setcounter{section}{0}

 \end{document}